\documentclass[sigconf]{acmart}
\AtBeginDocument{%
  }

\usepackage{xifthen, array, tabularx}
\usepackage{catchfile}

\usepackage{makecell, booktabs, geometry, url}%
\usepackage{hyperref}
\usepackage[utf8]{inputenc}
\usepackage{multirow}
\usepackage{enumitem}

\usepackage{float}
\floatstyle{plaintop}
\restylefloat{table}

\newcommand*{\MyPath}{./FIG}%

\newcommand{\test}[1][]{%
\ifthenelse{\equal{#1}{}}{omitted}{given}%
}

\IfFileExists{PAPERTAG.txt}
 {}
 {}

\graphicspath{{./FIG/}}

\newcommand{\MyFigRef}[1]{
Fig.~\ref{mypaper-fig#1}}

\usepackage{fancyvrb} 

\setcopyright{acmlicensed}
\copyrightyear{2025}
\acmYear{2025}
\acmDOI{XXXXXXX.XXXXXXX}
\acmConference[WCAE '25]{Workshop on Computer Architecture Education}{June 21--25,
  2025}{Tokyo, Japan}
\acmISBN{978-1-4503-XXXX-X/2025/06}

\begin{document}


\title[FREESS: An Educational Simulator of a  RISC-V-Inspired Superscalar Processor Based on Tomasulo’s Algorithm]{FREESS: An Educational Simulator of a RISC-V-Inspired Superscalar Processor Based on Tomasulo’s Algorithm}

\author{Roberto Giorgi}
\email{giorgi@unisi.it}
\orcid{0000-0003-0384-8229}

\affiliation{%
  \institution{University of Siena}
 \city{Siena}
 \country{Italy}
}

\renewcommand{\shortauthors}{Roberto Giorgi}

\begin{abstract}
FREESS is a free, interactive simulator that illustrates instruction-level parallelism in a RISC-V-inspired superscalar processor.
Based on an extended version of Tomasulo’s algorithm, FREESS is intended as a hands-on educational tool for Advanced Computer Architecture courses.
It enables students to explore dynamic, out-of-order instruction execution, emphasizing how instructions are issued as soon as their operands become available.

The simulator models key microarchitectural components, including the Instruction Window (IW), Reorder Buffer (ROB), Register Map (RM), Free Pool (FP), and Load/Store Queues.
FREESS allows users to dynamically configure runtime parameters, such as the superscalar issue width, functional unit types and latencies, and the sizes of architectural buffers and queues.

To simplify learning, the simulator uses a minimal instruction set inspired by RISC-V (ADD, ADDI, BEQ, BNE, LW, MUL, SW), which is sufficient to demonstrate key pipeline stages: fetch, register renaming, out-of-order dispatch, execution, completion, commit, speculative branching, and memory access.
FREESS includes three step-by-step, illustrated examples that visually demonstrate how multiple instructions can be issued and executed in parallel within a single cycle.
Being open source, FREESS encourages students and educators to experiment freely by writing and analyzing their own instruction-level programs and superscalar architectures.
\end{abstract}






\received{19 May 2025}

\maketitle

\section{Introduction}
\begin{figure}
\centering
\includegraphics[width=0.92\columnwidth]{\MyPath/\input{PAPERTAG.txt}-figsm_FIG_1.pdf}
\caption[]{\input{\MyPath/\input{PAPERTAG.txt}-figsm_Slide_1.txt}}
\Description{\input{\MyPath/\input{PAPERTAG.txt}-figsm_Slide_1.txt}}
\label{mypaper-fig1}
\end{figure}

Most modern microprocessors in medium- and high-end systems adopt \textit{superscalar} architectures, building upon dynamic scheduling mechanisms such as \textit{Thornton’s scoreboard}~\cite{Thornton70} and \textit{Tomasulo’s algorithm}~\cite{Tomasulo67}. In this paper, we focus on Tomasulo’s to illustrate how machine instructions can be executed \textit{in parallel}, transparently to the user. These techniques are commonly known as \textit{dynamic scheduling}, \textit{out-of-order execution}, or \textit{restricted dataflow}~\cite{Hwu86}.

This topic is therefore a cornerstone for \textit{Advanced Computer Architecture} courses, where \textit{Instruction-Level Parallelism (ILP)} plays a key role in achieving \textit{High-Performance Computing} and in accelerating \textit{single-threaded} execution.

Superscalar architectures provide an elegant hardware-based solution to the problem of tracking control and data dependencies among instructions.
While thread— and data-level parallelism are useful, executing multiple instructions per cycle can also significantly boost performance.
Normally, one instruction takes several cycles to complete. 
A first performance improvement comes from \textit{pipelining}, but a single pipeline can still issue at most one instruction per cycle, also known as \textit{Flynn's bottleneck}. 
This limitation is overcome by issuing multiple instructions to independent functional units. 
This goal is achieved in \textit{superscalar processors} (via hardware scheduling) or \textit{VLIW architectures} (via software scheduling).

\begin{figure*}[!htbp]
\centering
\includegraphics[width=0.7\textwidth]{\MyPath/\input{PAPERTAG.txt}-figsm_FIG_7.pdf}
\caption[]{\input{\MyPath/\input{PAPERTAG.txt}-figsm_Slide_7.txt}}
\Description{\input{\MyPath/\input{PAPERTAG.txt}-figsm_Slide_7.txt}}
\label{mypaper-fig7}
\end{figure*}

The superscalar approach based on Tomasulo's extended algorithm augments the pipeline with several hardware structures that dynamically track data dependencies and execute instructions according to the \textit{dataflow} principle: firing them as soon as their operands are ready. The most relevant structures involved in the tracking dependencies include: \textit{Physical Registers (Px)}, the \textit{Free Pool (FP)}, the \textit{Register Map (RM)}, the \textit{Instruction Window (IW)}, the \textit{Reorder Buffer (ROB)}, and the \textit{Load/Store Queues (LSQs)} (see \MyFigRef{1}).

Several superscalar simulators have been proposed to aid in teaching these concepts. These simulators effectively show the internal behavior of these hardware components. However, the information presented is often difficult to reproduce with pencil and paper, as updates are either overwritten on screen or dispersed across multiple disconnected windows.


\textbf{FREESS} (Free Educational Superscalar Simulator) addresses these limitations by providing a \textit{single screen} visualization summarizing both the current state and the cycle-by-cycle evolution of instructions through the superscalar pipeline.
This evolution matches exactly what can be written on a single sheet of paper (see \MyFigRef{7}) while solving one of the exercises used for student training and for preparation verification (exams). Moreover, the output of FREESS can be printed on paper for reviewing or studying purposes, even without the need to interact with a computer. This approach encourages students to bridge theoretical concepts with manual tracing and reinforces the broader course’s emphasis on performance-critical architectural design. 

The syntax of the code recalls RISC-V instructions and registers due to the popularity of this novel instruction set, which is not bound to a single manufacturer and is widely adopted in Computer Architecture classes. FREESS supports a small but sufficient set of seven instructions—\texttt{ADD}, \texttt{ADDI}, \texttt{BEQ}, \texttt{BNE}, \texttt{LW}, \texttt{MUL}, and \texttt{SW}—following a minimalistic approach in the spirit of other well-known educational tools like the LC3 simulator~\cite{Patt04-book}. Although adding more instructions is very simple in FREESS, the target of the simulator is studying the superscalar internals, not exploring a wide range of instructions, substitute debuggers, or other types of simulators.

\subsection*{Contributions}
The main contributions of this work are:
\begin{itemize}
\item to illustrate a methodology for teaching superscalar execution, where students can manually trace execution on paper using a layout that mirrors the one provided by the tool;
\item to present FREESS, a simulator that provides a cycle-accurate view of the key hardware structures used in a Tomasulo-based superscalar processor;
\item to support the open source release of FREESS, along with guidance on writing and visualizing RISC-V-like programs.
\end{itemize}

The rest of the paper is organized as follows: Section 2 discusses related work; Section 3 describes the simulator; Section 4 presents illustrative examples; Section 5 analyzes the educational impact; and Section 6 concludes the paper.

\section{Related Work}
Several educational simulators have been developed to support the teaching of superscalar architecture concepts, each offering various degrees of visualization and interactivity.

SIMDE \cite{castilla07-simde}] is designed to support dynamic and static scheduling through Tomasulo’s algorithm and scoreboarding. It enables students to explore the flow of instructions and the hazard resolution mechanisms, providing insight into out-of-order execution and register renaming. Although SIMDE effectively represents data hazards and scheduling logic, its visualization primarily focuses on instruction status tables and resource usage. It lacks a complete, unified cycle-by-cycle overview of the instruction pipeline.

SATSim \cite{Wolff00-satsim} provides an interactive, GUI-based environment to understand superscalar architectures. It emphasizes visual tracing of individual instructions and the state of internal buffers, which is valuable for observing execution behavior. However, SATSim offers limited feedback on stall causes or pipeline-wide performance trends, and it does not model the Load and Store Queues.

PSATSim \cite{Smullen06-psatsim} builds on SATSim by incorporating power and performance metrics. It allows users to explore the effects of microarchitectural changes on energy efficiency and throughput. Despite its improvements, PSATSim lacks comprehensive per-cycle visualization of pipeline stages.

Jaros \cite{Jaros24-riscv} introduces a Web-based RISC-V simulator with superscalar support. Although accessible and platform-independent, this simulator primarily focuses on instruction execution and register state. Its support for architectural configuration is limited and does not include detailed memory pipeline modeling or stall diagnostics.

Other simulators like Ripes \cite{Petersen21} and several others \cite{Mariotti22-softwarex} model RISC-V pipeline, but lack a detailed superscalar modeling.

In contrast, FREESS offers a more holistic approach by providing a cycle-by-cycle visualization of the full instruction pipeline, including the Instruction Window, Register Map, Reorder Buffer, and particularly the Load and Store Queues, which are often omitted in comparable tools. FREESS highlights instruction-level parallelism by showing how multiple instructions proceed concurrently through fetch, rename, issue, execute, complete, and commit stages. Crucially, it also reports stall conditions and their causes, such as structural, data, or control hazards, making it easier for students to understand pipeline bottlenecks and dependency resolution. This level of feedback helps bridge the gap between theoretical understanding and practical insight into superscalar processor behavior.

\begin{figure*}[!htbp]
\centering
\includegraphics[width=0.65\textwidth]{\MyPath/\input{PAPERTAG.txt}-figsm_FIG_6.pdf}
\caption[]{\input{\MyPath/\input{PAPERTAG.txt}-figsm_Slide_6.txt}}
\Description{\input{\MyPath/\input{PAPERTAG.txt}-figsm_Slide_6.txt}}
\label{mypaper-fig6}
\end{figure*}

\section{Description}
\subsection{Launching the simulator}
FREESS is written in pure C code, compatible with GCC from version 2.7 (year 2009) to present (year 2025) without a single warning. This characteristic makes the tool available on many computing platforms, including MS-Windows via the Ubuntu shell, for example, and, of course, all Linux-based systems. 
The tool is launched on the command line, and it first generates the text of the exercise, indicating the working hypothesis, based on the default options or the optional command parameters (\MyFigRef{6}). This feature is also useful to teachers for generating new exercises and for students to explore and recall the simulated architecture's key parameters. 
In \MyFigRef{2} and \MyFigRef{3}, some architectural parameters are shown on architectural sketch and listed with a more detailed description.
\begin{figure}[H]
\centering
\includegraphics[width=0.95\columnwidth]{\MyPath/\input{PAPERTAG.txt}-figsm_FIG_2.pdf}
\caption[]{\input{\MyPath/\input{PAPERTAG.txt}-figsm_Slide_2.txt}}
\Description{\input{\MyPath/\input{PAPERTAG.txt}-figsm_Slide_2.txt}}
\label{mypaper-fig2}
\end{figure}
\begin{figure}[H]
\centering
\includegraphics[width=0.9\columnwidth]{\MyPath/\input{PAPERTAG.txt}-figsm_FIG_3.pdf}
\caption[]{\input{\MyPath/\input{PAPERTAG.txt}-figsm_Slide_3.txt}}
\Description{\input{\MyPath/\input{PAPERTAG.txt}-figsm_Slide_3.txt}}
\label{mypaper-fig3}
\end{figure}

In \MyFigRef{7} from top to down, the state of the superscalar parameters can be located and monitored cycle-by-cycle: the first group indicates the register map (a star is appearing below the physical registers that are allocated), qi indicates if the register is free, and vi is the register content.
The second group indicates the logical registers (xi), the associated physical register (Pi), whether the value is actual (Qi), and its value (Vi).

The third group represents the accounting of the resources in terms of slots in the buffer stages (F, D, P, I, X, W, C)\footnote{The letters indicate the classic stages: F=Fetch, D=Decode/Renaming, P=Dispatch, I=Issue, X=Execute, W=Write-back and C=Commit.},
in the renaming logic, in the instruction window, in the reorder buffer, in each of the functional units (A, M, L, S, B, F, X)\footnote{The letters indicate the functional units: A=ALU, M=integer-multiplier, L=Load, S-Store, B=branch, F=floating-point add/sub, X=floating-point-mul/div.}.
The fourth group presents for each instruction: the dynamic program counter (PC), the cycle when the instruction enters a certain stage, and other info detailed in the paper's text and the next figures.

\subsection{Defining machine-code programs}
Currently, the simulator deliberately avoids the complexity of an assembly parser. The student can code a few assembly instructions manually instead. It is left to the open source community to pick up on this point and eventually add a more complete parser. 
Table~\ref{tab:instructions} reports the opcodes for the seven supported instructions.

To write a program, students simply replace each mnemonic with its corresponding opcode, specify the register indices, and provide any required immediate values. The handling of branch instructions is particularly instructive: instead of using labels as in assembly, students must enter the immediate value representing the number of instructions to jump—positive for forward branches, negative for backward ones, effectively replacing the role of labels in \texttt{BEQ} and \texttt{BNE} instructions. 

\begin{table}\small
  \caption{FREESS instructions}
  \label{tab:instructions}
  \begin{tabular}{cl}
    \toprule
    Mnemonic & Operation Code (opcode)\\
    \midrule
ADD & 1\\
ADDI &2 \\
LW &3 \\
SW &4 \\
BEQ &5 \\
BNE &6\\
MUL &7\\
  \bottomrule
\end{tabular}
\end{table}

\begin{figure}[H]
\centering
\includegraphics[width=0.76\columnwidth]{\MyPath/\input{PAPERTAG.txt}-figsm_FIG_4.pdf}
\caption[]{\input{\MyPath/\input{PAPERTAG.txt}-figsm_Slide_4.txt}}
\Description{\input{\MyPath/\input{PAPERTAG.txt}-figsm_Slide_4.txt}}
\label{mypaper-fig4}
\end{figure}
Assuming the code of \MyFigRef{4} the resulting code is the following:
\begin{center}\begin{minipage}{0.2\linewidth}
\ttfamily
1  3  4    0   \\
1  7  5  128 \\
7  7  7    3   \\
6  1  1   -1  \\
2  7  6  256 \\
6  2  2    8   \\
4  1  0   -7  \\
\end{minipage}\end{center}

As the processor can only execute machine code, one more bonus of FREESS is that it forces the student to realize that (or refreshing the concept if learned in a pre-requisite course) by manually converting the program from assembly mnemonic to a specific (simple) machine code format.

Superscalar execution can be analyzed in depth with assembly programs of just a few instructions (e.g., 5 to 10 instructions) and in a loop to see the effect of the branch speculation. Therefore, it is not difficult to code the program. As shown in the next Section, some pre-built examples are provided to simplify this task.

\subsection{Dynamic stream and status of superscalar hardware structures}
By default, the simulator assumes that the program executes three iterations of a loop (a command line parameter can change the number of iterations) and that the branch is speculatively assumed to be taken (a detailed branch predictor is future work).

\subsubsection{Fetch stage}

In \MyFigRef{7}, the output after the first cycle (cycle 0) is shown. The four vertical zeros under the 'F' indicate that the first four instructions are in the fetch stage.
The simulator executes the next cycle every time the Enter key is pressed.

\subsubsection{Decode/Rename stage}
The next cycle (cycle 1) is shown in \MyFigRef{8}. The next four instructions should go to the fetch stage. However, the branch forces the fetch stage to break fetching and get only three instructions instead of four. 
\begin{figure}[H]
\centering
\includegraphics[width=1.00\columnwidth]{\MyPath/\input{PAPERTAG.txt}-figsm_FIG_8.pdf}
\caption[]{\input{\MyPath/\input{PAPERTAG.txt}-figsm_Slide_8.txt}}
\Description{\input{\MyPath/\input{PAPERTAG.txt}-figsm_Slide_8.txt}}
\label{mypaper-fig8}
\end{figure}
The first four instructions progress and go to the decode/rename stage. 
The columns 'Pi Pj Pk Pl' report the renamed stream and show how the logical registers are renamed to physical registers. The destination register is taken from the free pool (if not available, there will be a structural hazard), and the Register Map is updated accordingly, instruction after instruction, but within the same cycle. 
The Register Map (RM) tracks the assignment of the source logical to physical registers and is always visible at the top of the screen.

Please note that the screen flows: the current screen presents a complete overview of the stream's evolution, and the previous screens are still available for double-checking and can be dumped into a file for further examination and study.

\subsubsection{Dispatch stage}
The dispatch stage receives instructions from the renaming stage. Depending on the implementation, it stores them into the structures that will hold them until they are ready to execute, i.e., the Instruction Window (IW) or Reservation Stations (RS). We call them just \textit{IW-SLOTS}. At the same time, a ROB entry is allocated. If either IW or ROB is full, we have a structural hazard. Again, the number of times this happens is annotated in the accounting area and explained at the bottom of the screen.

The IW-SLOT is a record that contains the entry identifier (IW\#), the opcode (OPCD), the destination register (Pi), the first source register (Pj), the second source register (Pk), and the immediate (I). To track the availability of the source register values, the "flags" Cj and Ck are used: here we extend their meaning: a positive number indicates the cycle when the corresponding physical register (j or k) received its value in the IW\footnote{A '-' means that the value is yet to be produced, so the instruction cannot be issued.}.

\begin{figure}[H]
\centering
\includegraphics[width=1.00\columnwidth]{\MyPath/\input{PAPERTAG.txt}-figsm_FIG_9.pdf}
\caption[]{\input{\MyPath/\input{PAPERTAG.txt}-figsm_Slide_9.txt}}
\Description{\input{\MyPath/\input{PAPERTAG.txt}-figsm_Slide_9.txt}}
\label{mypaper-fig9}
\end{figure}

In the example of \MyFigRef{9} instructions 0, 1, and 2 (first LW, second LW and ADDI) received their source value at cycle 2, i.e., when they entered the IW\footnote{We also introduced a possible third source register Pl and the corresponding flag Cl for future extensions.}.

When an IW-slot is allocated, a corresponding ROB-SLOT is. This can be easily identified on the screen since the IW-SLOT and the ROB-SLOT are on the same row, one beside the other.

The corresponding ROB-SLOT is a record that contains the entry identifier (ROB\#), the PC of the instruction, the destination logical register (xi) and eventually the old physical register (oPi) associated with that register in the Register Map (RM): this is useful when eventually there is an exception or miss-speculation and the execution has to be rolled-back.
There are also three flags: 's' to record whether the operation was associated with a store, 'x' to signal an exception, and 'c' to indicate when the instruction has completed, i.e., it has written back its result. \MyFigRef{9} shows that the IW and ROB are populated at cycle 2.
\begin{figure}[H]
\centering
\includegraphics[width=1.00\columnwidth]{\MyPath/\input{PAPERTAG.txt}-figsm_FIG_10.pdf}
\caption[]{\input{\MyPath/\input{PAPERTAG.txt}-figsm_Slide_10.txt}}
\Description{\input{\MyPath/\input{PAPERTAG.txt}-figsm_Slide_10.txt}}
\label{mypaper-fig10}
\end{figure}

\subsubsection{Issue stage}
Once all source values have reached the physical registers of the instructions in the IW, one or more instructions can be issued. 
The issue width could be restricted to a single instruction or multiple instructions, up to the issue width (another simulation parameter). The instructions are sent to the corresponding available Functional Units (FUs) in this phase. 

If the FU is occupied, we have a structural hazard, and the statistics on the screen are updated (see \MyFigRef{10}).
The issued instructions are also marked with the character '>' besides the IW\#. 
For example, in \MyFigRef{10}, the first LW and the ADDI are issued at cycle 3.

\paragraph{Load and Store Queues}

Loads or stores are queued in the load queue or store queue, respectively (indicated in the right part of the screen), along their PC, opcode (OP), and the Effective Address (EFAD), which is calculated before queuing, as shown in \MyFigRef{10}.
In case of a load, the value read from the memory hierarchy is written in Pi. Ci indicates the cycle when the reload is queued, and later it is updated with the cycle when the value is forwarded to the common data bus.
In the case of a store, the value to be written could yet to be produced, so Pl (and Cl) here indicate respectively the physical register waiting for such a value and the cycle when it arrives, which corresponds to the cycle when the store is ready to access memory in the order specified by the store queue (and in synchronization with the load queue).

\subsubsection{Execute stage}
During the execution stage (cycle 4 of this example), we can observe that some of the instructions are fired since the IW-SLOT identifier (IW\#) is becoming `- - - -' (see \MyFigRef{11}). The related information remains on the screen for reference, while the corresponding ROB-SLOT is freed only once the instruction commits.
Depending on the latency associated with the Functional Unit, the associated instruction continues to occupy the X stage for the corresponding number of cycles, assuming the unit is pipelined. In the figure \MyFigRef{1}, the multiplication consists of 4 stage (X1, X2, X3, X4) for the sake of exemplification.

For ALU operations and the EFAD calculation, the Issue (I) and Execute (X) stages often operate on the same cycle: this is our default behavior. However, as a simulation parameter, the user can also specify that I and X should always happen in different cycles.

\subsubsection{Write-back (or Complete) stage}
In our driving example at cycle 4, the ADDI operation completes, as can be seen, out-of-order (see \MyFigRef{11}). This is marked by c=1. However, the ROB-SLOT cannot be freed until all previous ROB-SLOTs have the c flag equal to 1, ensuring that the logical registers are updated in program order. The ROB is managed as a circular queue. In case of exceptions or mis-speculations, the operations must be undone by properly updating the Register Map and the Free Pool. Several techniques exist, but are beyond our scope here.
\begin{figure}[H]
\centering
\includegraphics[width=1.00\columnwidth]{\MyPath/\input{PAPERTAG.txt}-figsm_FIG_11.pdf}
\caption[]{\input{\MyPath/\input{PAPERTAG.txt}-figsm_Slide_11.txt}}
\Description{\input{\MyPath/\input{PAPERTAG.txt}-figsm_Slide_11.txt}}
\label{mypaper-fig11}
\end{figure}

\subsubsection{Commit stage}
The student can observe how the execution progresses by observing the cycles in which each instruction enters a pipeline stage and the updates in the RM, FP, IW, ROB, LQ, and SQ. Once some instructions - up to the commit width (4 or fewer, in our case) - are completed \textit{and} are on top of the circular queue, then those instructions are committed.

At the end of the simulation (\MyFigRef{12}), the whole evolution of the execution and the final statistics are still visible on the screen. In particular, the driving example has taken 20 cycles (CTOT) and the Instruction-Per-Cycle (IPC) is 1.05, meaning that the pipeline has been able to achieve, on average, slightly more than one instruction per cycle, despite the data dependencies and the FU latencies.
\begin{figure*}[!htbp]
\centering
\includegraphics[width=0.89\textwidth]{\MyPath/\input{PAPERTAG.txt}-figsm_FIG_12.pdf}
\caption[]{\input{\MyPath/\input{PAPERTAG.txt}-figsm_Slide_12.txt}}
\Description{\input{\MyPath/\input{PAPERTAG.txt}-figsm_Slide_12.txt}}
\label{mypaper-fig12}
\end{figure*}

\section{Examples}
The FREESS package has three examples, the first described in detail in the previous section. For easier start up, the examples can be launched via a script \texttt{./run-exK.sh}, where \texttt{K} is '1', '2', or '3'. Once the student is more accustomed to the FREESS workflow, they can write her/his own programs (or modify the scripts), study the effect of different superscalar architectural parameters by sampling them on the command line as shown in \MyFigRef{28}.
\begin{figure}[H]
\centering
\includegraphics[width=0.95\columnwidth]{\MyPath/\input{PAPERTAG.txt}-figsm_FIG_28.pdf}
\caption[]{\input{\MyPath/\input{PAPERTAG.txt}-figsm_Slide_28.txt}}
\Description{\input{\MyPath/\input{PAPERTAG.txt}-figsm_Slide_28.txt}}
\label{mypaper-fig28}
\end{figure}

\subsection{Example-2}
The second example is a smaller loop consisting of only five instructions. \MyFigRef{13} shows the auto-generated text of this example, and the green box highlights the major modifications of the architecture: 12 IW-SLOTS, 12 Physical Registers, 12 ROB-SLOTS.
\begin{figure}[H]
\centering
\includegraphics[width=1.0\columnwidth]{\MyPath/\input{PAPERTAG.txt}-figsm_FIG_13.pdf}
\caption[]{\input{\MyPath/\input{PAPERTAG.txt}-figsm_Slide_13.txt}}
\Description{\input{\MyPath/\input{PAPERTAG.txt}-figsm_Slide_13.txt}}
\label{mypaper-fig13}
\end{figure}

As can be seen in the final output windows, in this case, the superscalar can achieve an IPC of 1.36 with 11 cycles of execution. This means that several more instructions are now run in parallel.
\begin{figure}[H]
\centering
\includegraphics[width=1.00\columnwidth]{\MyPath/\input{PAPERTAG.txt}-figsm_FIG_18.pdf}
\caption[]{\input{\MyPath/\input{PAPERTAG.txt}-figsm_Slide_18.txt}}
\Description{\input{\MyPath/\input{PAPERTAG.txt}-figsm_Slide_18.txt}}
\label{mypaper-fig18}
\end{figure}

Comparing \MyFigRef{13} and \MyFigRef{18}, we can observe that the second program generates fewer stalls due to lacking resources. In the first example (\MyFigRef{13}), we got 9 stalls due to renaming (D stage), 12 due to dispatch (P stage), 16 due to Issue (I stage), and 7 due to commit (C stage). In the second examples (\MyFigRef{18}), these numbers are respectively 0 (D), 0 (P), 6 (I), 0(C)\footnote{We are reporting only those statistics that are relevant for the comparison with the previous example.}.

\subsection{Example-3}
To confirm the conclusions of Example-2, another slightly different example is considered: Example-3. In this case, the program remains the same as in the Example-2, but the superscalar width is reduced to simulate a 2-way superscalar. We omit the auto-generated text to save some space, and we report the final statistics of the execution in \MyFigRef{24}. In this case, the stalls are 0(D), 3 (P), 3 (I), 0 (C). While the total number of stalls is the same (6) in example-2 and example-3, the ability to process and commit fewer instructions per cycle is limiting the IPC to 1.07 with 14 total executed cycles.
\begin{figure}[H]
\centering
\includegraphics[width=1.00\columnwidth]{\MyPath/\input{PAPERTAG.txt}-figsm_FIG_24.pdf}
\caption[]{\input{\MyPath/\input{PAPERTAG.txt}-figsm_Slide_24.txt}}
\Description{\input{\MyPath/\input{PAPERTAG.txt}-figsm_Slide_24.txt}}
\label{mypaper-fig24}
\end{figure}

\subsection{Analyzing the Stall Reasons}
While the stall explanation is reported in the bottom part of the screen cycle-by-cycle (see \MyFigRef{10} and \MyFigRef{11}), the same entries (all stalls) are also logged in the \texttt{stall.log} file for final reference (see \MyFigRef{30}). Each stall explanation reports: the cycle when the stall happened, the reason for the stall, the involved instruction, and the stage where this happened. 
By analyzing these explanations, students gain deeper insight into whether stalls are caused by resource constraints or particular instructions that could be optimized in the source program. Experimenting with architectural parameters, for example in Example-3, they can see how increasing the dispatch width, issue width, and the number of ALUs (e.g., using \texttt{-pw 3 -iw 3 -afu 2}) eliminates dispatch-stage stalls. Comparing across different configurations (such as 2-way versus 4-way superscalar) further highlights how architectural decisions directly affect performance. This exercise provides students with a hands-on understanding of how hardware limitations influence stalls and hence performance.
\begin{figure}[H]
\centering
\includegraphics[width=1.00\columnwidth]{\MyPath/\input{PAPERTAG.txt}-figsm_FIG_30.pdf}
\caption[]{\input{\MyPath/\input{PAPERTAG.txt}-figsm_Slide_30.txt}}
\Description{\input{\MyPath/\input{PAPERTAG.txt}-figsm_Slide_30.txt}}
\label{mypaper-fig30}
\end{figure}

\section{Impact}
Superscalar processors and out-of-order execution are key topics in modern computer architecture courses, forming the foundation of high-performance CPU design. These topics are commonly covered in graduate curricula using textbooks such as \textit{Computer Architecture: A Quantitative Approach} by Hennessy and Patterson~\cite{Hennessy17} and \textit{Parallel Computer Organization and Design} by Dubois, Annavaram, Stenstr\"{o}m~\cite{Dubois12}. However, while superscalar execution is often supported by visual tools and exercises, the more advanced concepts of superscalar execution, register renaming, and instruction reordering are harder to teach and visualize.

FREESS (Free Educational Superscalar Simulator) addresses this pedagogical gap by offering a lightweight, open-source tool to support teaching Tomasulo-style, out-of-order superscalar execution. It provides a clear, cycle-by-cycle visualization of how instructions move through the pipeline—from fetch to commit, while showing internal structures such as the Instruction Window (IW), Reorder Buffer (ROB), Register Map (RM), Free Pool (FP), and Load/Store Queues (LSQs). FREESS displays the current state of each instruction and hardware structure and highlights the causes and frequency of stalls (e.g., structural, data, and control hazards), offering a practical and detailed perspective on pipeline bottlenecks.

The RISC-V community has grown exponentially, as has the demand for more performance in RISC-V implementations for High-Performance Computing. Therefore, several RISC-V-based efforts worldwide are exploring superscalar designs to improve the execution performance.
Using a RISC-V-inspired instruction set, FREESS aligns with the increasing shift in academia from older ISAs like MIPS toward the open and modern RISC-V standard. The simplified encoding of instructions and manual entry of opcode and register indices helps students understand the fundamentals of instruction encoding and control flow.

FREESS has been used in the Computer Architecture curriculum at our institution since 2010 and is distributed along with illustrative examples and configuration scripts. It enables students to engage with realistic yet manageable exercises that reflect the behavior of real-world superscalar processors. The tool supports active learning, allowing students to replicate the simulation outputs on paper for deepened understanding. About 4 hours of lessons and 6 hours of practicing are planned for teaching dynamic scheduling and superscalar concept (plus 2 hours on branch prediction) at the University of Siena (course site: \url{https://hpca.dii.unisi.it/} ). The related slides used for the teaching are available on demand.

In our experience, students initially struggle to understand out-of-order execution, but after using FREESS to manually step through cycle-by-cycle outputs, they consistently report feeling more confident in tracing out-of-order execution. They appreciate how the textual interface matches what they do in paper-based exercises, and in seconds, they can evaluate the effectiveness of different architectural choices in the structure of the superscalar.

FREESS also has limitations in that it does not model everything currently available in the current superscalar processor, but only the main structures typically addressed at the level of an advanced course in Computer Architecture. FREESS is not a production tool, meaning there is a lot of space for improving it for different needs. The fact that it is written in pure C and has a very limited complexity of about 2000 lines of code (including lots of comments and pretty-printing functions) should make its extension easy to perform. 

Finally, FREESS open-source nature encourages contributions and extensions. We envision a growing community of educators and students around FREESS, sharing new exercises, architectural variants, and features, further enhancing the tool’s utility and educational reach.

\section{Conclusions}
FREESS (Free Educational Superscalar Simulator) provides an effective and accessible tool for teaching the principles of superscalar processors and out-of-order execution, which are fundamental to modern computer architecture education. By offering a cycle-accurate visualization of key hardware structures such as the Instruction Window (IW), Reorder Buffer (ROB), Register Map (RM), Free Pool (FP), and Load/Store Queues (LSQs), FREESS bridges the gap between theoretical concepts and practical understanding. Its unified, text-based interface allows students to manually trace execution steps, reinforcing their comprehension of dynamic scheduling and dependency resolution.

The simulator's minimalistic RISC-V-inspired instruction set simplifies learning while maintaining relevance to contemporary architectures. FREESS ability to dynamically configure architectural parameters and log stall conditions enables students to explore the impact of resource limitations and pipeline hazards, fostering deeper insights into performance bottlenecks. Including illustrative examples and open-source availability further enhances its utility as an educational resource.

By abstracting away unnecessary complexity and focusing on a minimal but representative instruction set, FREESS lowers the barrier for understanding key pipeline stages while maintaining technical accuracy. Its visual and interactive approach encourages engagement, experimentation, and a deeper conceptual grasp of modern processor design. Its lightweight implementation in pure C ensures broad compatibility and ease of extension, inviting contributions from the educational community. 

Avoiding the use of a GUI has the further advantage of simply printing the output of the whole evolution of the execution on a cycle-by-cycle basis or an interesting part of it, for off-line study.

Future work includes extending FREESS with additional RISC-V instructions and developing more sophisticated branch prediction strategies. These enhancements will further enhance its value as both a teaching aid and a platform for architectural prototyping.

Another future possibility is to convert the C-code to a WebAssembly version for a Web-based execution.

In summary, FREESS addresses the challenges of teaching superscalar execution by providing a hands-on, interactive, and adaptable platform. Its open-source nature and alignment with RISC-V position it as a valuable resource for educators and students, promoting active learning and experimentation in Computer Architecture.

\begin{acks}
I want to thank the anonymous reviewers for their encouraging comments and Jonnatan Mendoza of BSC for the useful feedback.
The European Commission partially supported this work under the projects: AXIOM H2020 (id. 645496), TERAFLUX (id. 249013), HiPEAC (id. 101069836)
and EDGE-ME under the Next Generation EU via the Italian National Recovery and Resilience
Plan M4C2-Inv.1.4, CUP J33C22001170001, (CN00000013 - partnership ICSC).
\end{acks}

\bibliographystyle{ACM-Reference-Format}
\bibliography{refs}


\begin{thebibliography}{12}


\ifx \showCODEN    \undefined \def \showCODEN     #1{\unskip}     \fi
\ifx \showISBNx    \undefined \def \showISBNx     #1{\unskip}     \fi
\ifx \showISBNxiii \undefined \def \showISBNxiii  #1{\unskip}     \fi
\ifx \showISSN     \undefined \def \showISSN      #1{\unskip}     \fi
\ifx \showLCCN     \undefined \def \showLCCN      #1{\unskip}     \fi
\ifx \shownote     \undefined \def \shownote      #1{#1}          \fi
\ifx \showarticletitle \undefined \def \showarticletitle #1{#1}   \fi
\ifx \showURL      \undefined \def \showURL       {\relax}        \fi
\providecommand\bibfield[2]{#2}
\providecommand\bibinfo[2]{#2}
\providecommand\natexlab[1]{#1}
\providecommand\showeprint[2][]{arXiv:#2}

\bibitem[Castilla et~al\mbox{.}(2007)]%
        {castilla07-simde}
\bibfield{author}{\bibinfo{person}{I. Castilla}, \bibinfo{person}{L. Moreno},
  \bibinfo{person}{C. González}, \bibinfo{person}{J. Sigut}, {and}
  \bibinfo{person}{E. González}.} \bibinfo{year}{2007}\natexlab{}.
\newblock \showarticletitle{SIMDE: An Educational Simulator of ILP
  Architectures with Dynamic and Static Scheduling}.
\newblock \bibinfo{journal}{\emph{Comp. App. in Eng. Education}}
  \bibinfo{volume}{15}, \bibinfo{number}{4} (\bibinfo{year}{2007}),
  \bibinfo{pages}{309--318}.
\newblock
\href{https://doi.org/10.1002/cae.20154}{doi:\nolinkurl{10.1002/cae.20154}}


\bibitem[Dubois et~al\mbox{.}(2012)]%
        {Dubois12}
\bibfield{author}{\bibinfo{person}{Michel Dubois}, \bibinfo{person}{Murali
  Annavaram}, {and} \bibinfo{person}{Per Stenström}.}
  \bibinfo{year}{2012}\natexlab{}.
\newblock \bibinfo{booktitle}{\emph{Parallel Computer Organization and
  Design}}.
\newblock \bibinfo{publisher}{Cambridge University Press},
  \bibinfo{address}{Cambridge}.
\newblock


\bibitem[Hennessy and Patterson(2017)]%
        {Hennessy17}
\bibfield{author}{\bibinfo{person}{John~L. Hennessy} {and}
  \bibinfo{person}{David~A. Patterson}.} \bibinfo{year}{2017}\natexlab{}.
\newblock \bibinfo{booktitle}{\emph{Computer Architecture, Sixth Edition: A
  Quantitative Approach} (\bibinfo{edition}{6th} ed.)}.
\newblock \bibinfo{publisher}{MKP Inc.}, \bibinfo{address}{San Francisco, CA,
  USA}.
\newblock
\showISBNx{0128119055}


\bibitem[Hwu and Patt(1986)]%
        {Hwu86}
\bibfield{author}{\bibinfo{person}{W. Hwu} {and} \bibinfo{person}{Y.~N. Patt}.}
  \bibinfo{year}{1986}\natexlab{}.
\newblock \showarticletitle{HPSm, a high performance restricted data flow
  architecture having minimal functionality}.
\newblock \bibinfo{journal}{\emph{SIGARCH Comput. Archit. News}}
  \bibinfo{volume}{14}, \bibinfo{number}{2} (\bibinfo{date}{May}
  \bibinfo{year}{1986}), \bibinfo{pages}{297–306}.
\newblock
\showISSN{0163-5964}
\href{https://doi.org/10.1145/17356.17391}{doi:\nolinkurl{10.1145/17356.17391}}


\bibitem[Jaros(2024)]%
        {Jaros24-riscv}
\bibfield{author}{\bibinfo{person}{J. Jaros}.} \bibinfo{year}{2024}\natexlab{}.
\newblock \showarticletitle{Web-Based Simulator of Superscalar RISC-V
  Processors}. In \bibinfo{booktitle}{\emph{ICS'24}}.
  \bibinfo{publisher}{IEEE}, \bibinfo{address}{Piscataway, NJ, USA},
  \bibinfo{pages}{1--6}.
\newblock
\href{https://doi.org/10.1109/SCW63240.2024.00209}{doi:\nolinkurl{10.1109/SCW63240.2024.00209}}


\bibitem[Mariotti and Giorgi(2022)]%
        {Mariotti22-softwarex}
\bibfield{author}{\bibinfo{person}{G. Mariotti} {and} \bibinfo{person}{R.
  Giorgi}.} \bibinfo{year}{2022}\natexlab{}.
\newblock \showarticletitle{WebRISC-V: A 32/64-bit RISC-V pipeline simulation
  tool}.
\newblock \bibinfo{journal}{\emph{ELSEVIER SoftwareX}}  \bibinfo{volume}{18}
  (\bibinfo{date}{May} \bibinfo{year}{2022}), \bibinfo{pages}{1--7}.
\newblock
\showISSN{2352-7110}
\href{https://doi.org/10.1016/j.softx.2022.101105}{doi:\nolinkurl{10.1016/j.softx.2022.101105}}


\bibitem[Patt and Patel(2004)]%
        {Patt04-book}
\bibfield{author}{\bibinfo{person}{Yale Patt} {and} \bibinfo{person}{Sanjay
  Patel}.} \bibinfo{year}{2004}\natexlab{}.
\newblock \bibinfo{booktitle}{\emph{Introduction to Computing Systems: From
  Bits and Gates to C and Beyond} (\bibinfo{edition}{2} ed.)}.
\newblock \bibinfo{publisher}{McGraw-Hill}, \bibinfo{address}{New York, NY,
  USA}.
\newblock
\showISBNx{978-0-07-246750-5}


\bibitem[Petersen(2021)]%
        {Petersen21}
\bibfield{author}{\bibinfo{person}{Morten~B. Petersen}.}
  \bibinfo{year}{2021}\natexlab{}.
\newblock \showarticletitle{Ripes: A Visual Computer Architecture Simulator}.
  In \bibinfo{booktitle}{\emph{ISCA-WCAE'21}}. \bibinfo{publisher}{IEEE},
  \bibinfo{address}{Virtual Conference}, \bibinfo{pages}{1--8}.
\newblock


\bibitem[Smullen(2006)]%
        {Smullen06-psatsim}
\bibfield{author}{\bibinfo{person}{C.~W. Smullen}.}
  \bibinfo{year}{2006}\natexlab{}.
\newblock \showarticletitle{PSATSim: An Interactive Graphical Superscalar
  Architecture Simulator for Power and Performance Analysis}. In
  \bibinfo{booktitle}{\emph{ISCA-WCAE'06}}. \bibinfo{publisher}{ACM},
  \bibinfo{address}{New York, NY, USA}, \bibinfo{pages}{1--6}.
\newblock
\href{https://doi.org/10.1145/1275620.1275627}{doi:\nolinkurl{10.1145/1275620.1275627}}


\bibitem[Thornton({[n.\,d.]})]%
        {Thornton70}
\bibfield{author}{\bibinfo{person}{James~E. Thornton}.}
  \bibinfo{year}{[n.\,d.]}\natexlab{}.
\newblock \bibinfo{booktitle}{\emph{Design of a Computer: The Control Data
  6600}}.
\newblock \bibinfo{publisher}{Scott, Foresman and Co.},
  \bibinfo{address}{Glenview, IL, USA}.
\newblock


\bibitem[Tomasulo(1967)]%
        {Tomasulo67}
\bibfield{author}{\bibinfo{person}{Robert~M. Tomasulo}.}
  \bibinfo{year}{1967}\natexlab{}.
\newblock \showarticletitle{An Efficient Algorithm for Exploiting Multiple
  Arithmetic Units}.
\newblock \bibinfo{journal}{\emph{IBM Journal of Research and Development}}
  \bibinfo{volume}{11}, \bibinfo{number}{1} (\bibinfo{year}{1967}),
  \bibinfo{pages}{25--33}.
\newblock


\bibitem[Wolff(2000)]%
        {Wolff00-satsim}
\bibfield{author}{\bibinfo{person}{S. Wolff}.} \bibinfo{year}{2000}\natexlab{}.
\newblock \showarticletitle{SATSim: A Superscalar Architecture Trace Simulator
  Using Interactive Visualization}. In
  \bibinfo{booktitle}{\emph{ISCA-WCAE'00}}. \bibinfo{publisher}{ACM},
  \bibinfo{address}{New York, NY, USA}, \bibinfo{pages}{1--7}.
\newblock
\href{https://doi.org/10.1145/1275240.1275249}{doi:\nolinkurl{10.1145/1275240.1275249}}


\end{thebibliography}

\appendix
\section{Online Resources}
The FREESS Educational Simulator for RISC-V inspired Superscalar Processors based on Tomasulo's Algorithm is available at this address:
\textbf{\url{https://github.com/robgiorgi/freess}}
\end{document}